\documentclass[twocolumn,prl,superscriptaddress]{revtex4}
\usepackage[utf8]{inputenc}
\usepackage{amsmath}
\usepackage{amssymb}
\usepackage{bm}
\usepackage{graphicx}
\usepackage{color}

\begin{document}

\title{Insulating regime of an underdamped current-biased Josephson junction \\ supporting $\mathbb{Z}_3$ and $\mathbb{Z}_4$ parafermions}
\author{Aleksandr E. Svetogorov}
\author{Daniel Loss}
\author{Jelena Klinovaja}
\affiliation{University of Basel, Department of Physics, Klingelbergstrasse, 4056 Basel, Switzerland}

\date{\today}
\begin{abstract}
We study analytically a current-biased topological Josephson junction supporting $\mathbb{Z}_n$ parafermions. First, we show that in an infinite-size system a pair of parafermions on the junction can be in $n$ different states; the $2\pi{n}$ periodicity of the phase potential of the junction results in a significant suppression of the maximal current $I_m$ for an insulating regime of the underdamped junction. Second, we study the behaviour of a realistic finite-size system with avoided level crossings characterized by splitting $\delta$. We consider two limiting cases: when the phase evolution may be considered adiabatic, which results in decreased periodicity of the effective potential, and the opposite case, when Landau-Zener transitions restore the $2\pi{n}$ periodicity of the phase potential. The resulting current $I_m$ is exponentially different in the opposite limits, which allows us to propose a new detection method to establish the appearance of parafermions in the system experimentally, based on measuring $I_m$ at different values of the splitting $\delta$.
\end{abstract}
\maketitle

\begin{figure}
\includegraphics[width=7cm]{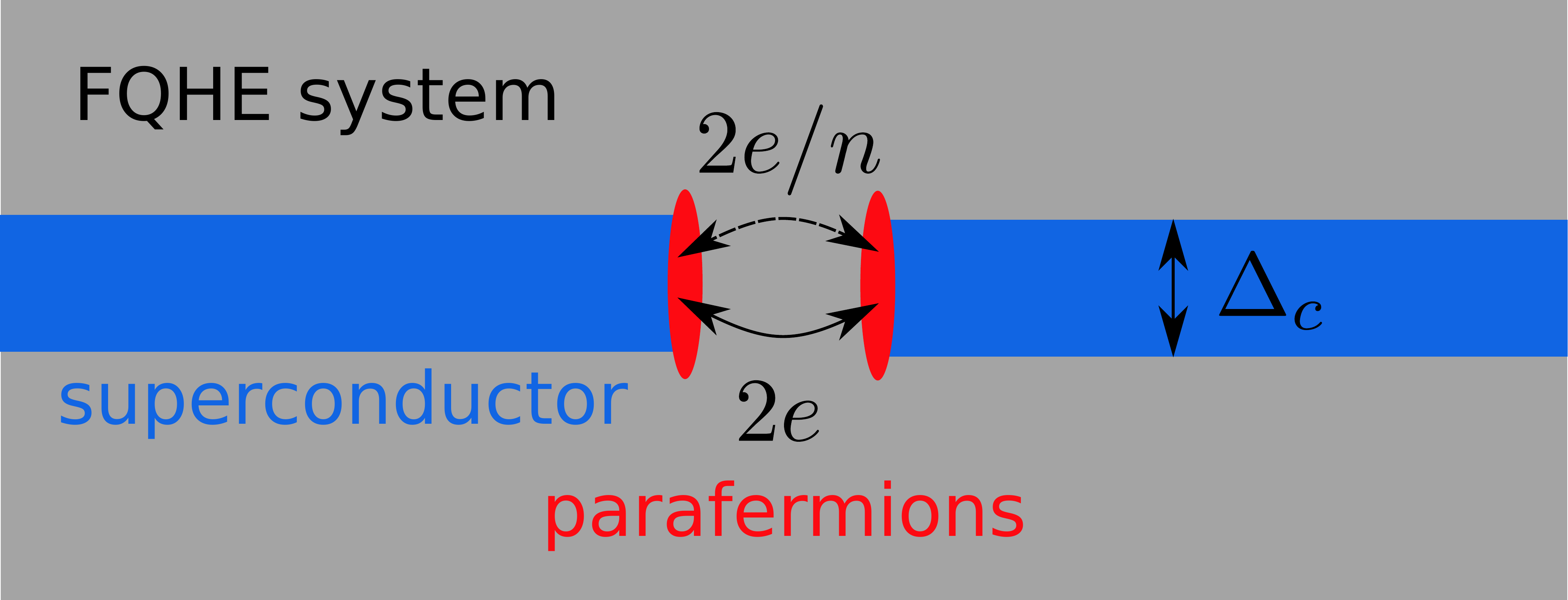}
\caption{\label{fig:structure}
Schematic representation of the FQHE stucture. Narrow superconducting strips (blue) induce pairing of amplitude $\Delta_c$ between counterpropagating FQHE edge states. Two strips placed close to each other form an effective JJ, a pair of $\mathbb{Z}_n$ parafermions on the junction forms a channel for $2e/n$ fractional quasiparticles tunneling between the superconducting strips along with ordinary Cooper pairs of charge $2e$.}
\end{figure}

{\it Introduction.}
Topological superconducting systems have recently attracted much attention both from a fundamental point of view and as possible platforms of a quantum computer~\cite{Kitaev2001,Nayak2008,Alicea2011}. One of the effects, which may indicate the topological properties of the system, is the fractional Josephson effect~\cite{Kitaev2001,Kwon2003,Kane2009,Wiedenmann2016}. In a trivial Josephson junction (JJ) the low-energy properties (like Josephson current) are determined by a $2\pi$ periodic phase potential. The best studied fractional Josephson effect is in junctions formed by topological superconductors supporting Majorana bound states (MBSs)~\cite{Meyer2011,Oppen2011,Furdyna2012,Alicea2014,Prada2012}. In this case, the phase potential of the system is $4\pi$ periodic due to the possibility of coherent transfer of a single electron as a result of the coupling of the MBSs on the sides of the junction. However, it is known that quasiparticle poisoning can spoil the $4\pi$ periodicity, which is a potential problem for all systems hosting MBSs~\cite{Kane2009,Chamon2011,Beenakker2011,Rainis2012,Trauzettel2012}.
Moreover, MBSs have Ising type  braiding statistics, which is not sufficient for universal quantum computation~\cite{Alicea2011,Oppen2012}. More exotic effects are predicted for systems with $\mathbb{Z}_n$ symmetries ($n>2$), the domain walls between topological and trivial phases host $\mathbb{Z}_n$ parafermions~\cite{Fendley2012,Kane2014,Yakoby2014,Loss2014,Fisher2014,Stern2014,Schmidt2015,Klinovaja2014,Klinovaja2015,
Fendley2016,Beri2016,Oreg2017,Schmidt2017,Alicea2018,Wu2018,Mora2018,Laubscher2019,Trauzettel2019} 
with more complex braiding statistics, which allows one to perform an entangling gate and makes parafermions computationally more powerful than MBSs~\cite{Clarke2013,Loss2016}. The effective state formed by a pair of parafermions carries fractional charge $2e/n$, which is robust against extrinsic quasiparticles (integer-charge quasiparticles cannot induce transitions between the $n$ possible states of the system). In general, the emergence of parafermions is predicted for  systems with strong electron-electron interactions; a pair of parafermions on the junction sides enables the tunneling of $2e/n$ fractional quasiparticles and, therefore, results in a $2\pi{n}$ periodicity in the phase~\cite{Cheng2012,Stern2012,Clarke2013,Kane2014,Loss2014,Lutchyn2015,Schmidt2015,Oppen2016,Schmidt2017}. The Hamiltonian of such a system takes the form
\begin{equation}
H=\frac{q^2}{2C}+U(\phi),\quad\left[\phi,q\right]=2ei,
\end{equation}
where the first term corresponds to the charging energy: $C$ is the capacitance of the junction, $q$ is the charge on the junction; $U(\phi)$ is the $2\pi{n}$-periodic phase potential.
 An experimental demonstration of parafermion edge states presents a complex problem. However, recent experiments on induced superconductivity in edge states of systems with fractional quantum Hall effect (FQHE) seem to be promising for this purpose~\cite{Herrero2016,Kim2017,Umansky2018,Kim2020}. A crossed Andreev pairing gap, $\Delta_c$, across the superconductor separating two counter-propagating edge states has been reported~\cite{Kim2017,Kim2020}, which is supposed to be sufficient for the formation of parafermions~\cite{Clarke2013}. We propose that for the direct observation of parafermions one needs to combine two such setups  into an effective JJ (see Fig.~\ref{fig:structure}), so that the fractional Josephson effect can be observed. Moreover, we discuss a general experimental method, which first has been introduced for JJs hosting MBSs~\cite{Svetogorov2020}, to distinguish topological junctions hosting parafermions from non-topological junctions, based on the properties of an underdamped JJ, which depends crucially on the periodicity of the phase potential.

We start with a general model of a JJ hosting $\mathbb{Z}_3$ or $\mathbb{Z}_4$ parafermions on the junction sides. We discuss the voltage peak $V_m=RI_m$ in the $I$-$V$ characteristics of such a device, shunted by a large resistance $R$ and biased by a current. This peak corresponds to a transition from an effectively insulating to a conducting state~\cite{Likharev1985,Ioffe2007,Zazunov2008}; its magnitude depends on the tunneling amplitude between the minima of the phase potential, therefore, the $2\pi{n}$ periodicity plays a curial role in this effect. Moreover, if one can control the transitions between the $n$ possible states of a parafermion pair on the junction (i.e. tuning the splitting $\delta$ at avoided crossings by changing the chemical potential~\cite{Burnell2016}), one can effectively change the periodicity of the potential and, as a result, control the value of $V_m$.
In our work, we consider the temperature to be low enough, i.e. $T\ll\omega_0$, with $\omega_0$ being the level spacing in the minima of the phase potential $U(\phi)$, to ignore  thermal fluctuations, which, in general, would result in smoothening of the voltage peaks.

\begin{figure}
\includegraphics[width=7cm]{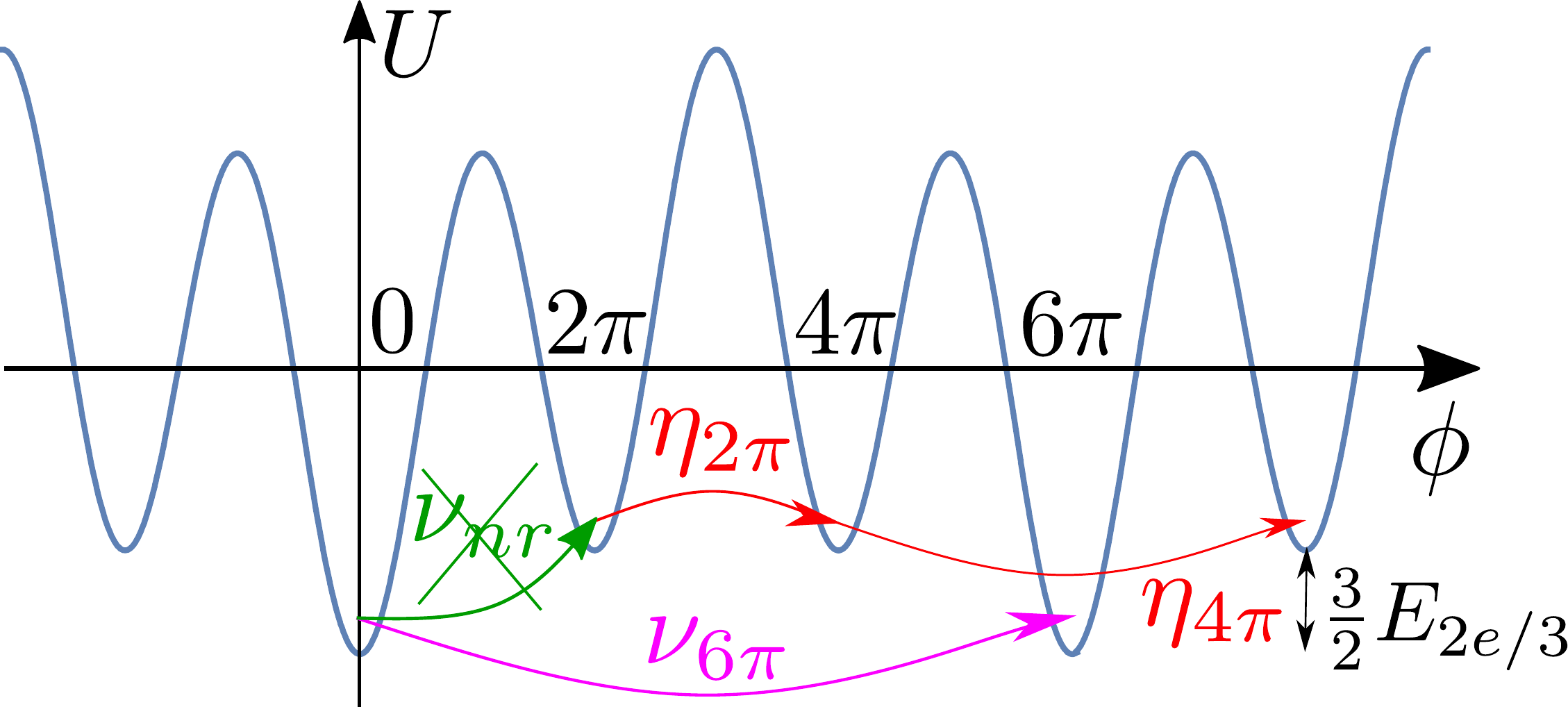}
\caption{\label{fig:6pi}
The phase potential $U$ [see Eq.~(\ref{phasepotential}) with $m=0$] of the junction supporting $\mathbb{Z}_3$ parafermions. The lowest band is determined by $6\pi$ tunneling with amplitude $\nu_{6\pi}$, while the non-resonant tunneling amplitude $\nu_{nr}$ is suppressed due to the energy shift of the next local minima by $3E_{2e/3}/2$. The next band is determined by $2\pi$ and $4\pi$ tunneling with amplitudes $\eta_{2\pi}$ and $\eta_{4\pi}$, respectively. }
\end{figure}

{\it $\mathbb{Z}_3$ case.} A pair of $\mathbb{Z}_3$ parafermions coupled via a JJ allows the transport of $2e/3$ fractional quasiparticles through the junction. As a result, the phase potential of the junction takes the form~\cite{Loss2014,Fisher2014,Fendley2016} 
\begin{equation}
U(\phi)=-E_J\cos\phi-E_{2e/3}\cos\left(\frac{\phi-2\pi{m}}{3}\right),
\label{phasepotential}
\end{equation}
 where $E_{2e/3}$ is the parafermion coupling amplitude, which governs fractional quasiparticles tunneling; $m \in \{0,1,2\}$ corresponds to one of the three states of the tunnel-coupled parafermion pair; $E_J$ corresponds to Cooper-pair tunneling through the junction. We consider the regime of a well-defined phase, i.e. $E_{c}=\frac{e^{2}}{2C}\ll E_{J}$, and  we assume the trivial Josephson tunneling to be dominant, $E_{J}\gg E_{2e/3}$. The lowest band depends only on $6\pi$ tunneling. We note that $2\pi$ tunneling is non-resonant, its amplitude $\nu_{nr}$ is smaller than the energy difference between neighboring
 local minima separated by $2\pi$: $\nu_{nr}\ll E_{2e/3}$. Therefore, this non-resonant tunneling is suppressed, see Fig.~\ref{fig:6pi}. Then, the lowest
energy band dispersion takes the form~\cite{Likharev1985,Ioffe2007}
\begin{equation}
E^{(0)}\left(k\right)=\frac{\omega_{0}}{2}-2\nu_{6\pi}\cos\left(6\pi k\right),
\end{equation}
where $
\omega_{0}\approx\sqrt{8E_{J}E_{c}}\left(1+\frac{1}{18}\frac{E_{2e/3}}{E_{J}}\right)$
is the harmonic frequency for the low-energy bands, $\nu_{6\pi}$ is the amplitude for $6\pi$ tunneling between the ground states in the absolute minima of $U(\phi)$.
It is convenient to compare it to the amplitude of $2\pi$ tunneling in a trivial junction~\cite{Likharev1985,Matveev2002}:
\begin{equation}\label{eq:nu_0}
\nu_0=\frac{4E_c}{\sqrt{\pi}}\left(\frac{2E_J}{E_c}\right)^{3/4}e^{-S_{0}},\, S_{0}=\sqrt{8E_J/E_c}. 
\end{equation}  We can calculate an instanton action for a $6\pi$ phase slip in a topological junction (expansion in $E_{2e/3}/E_J$):
\begin{equation}
S_{6\pi}
=3S_{0}\left(1+\left[1+\ln\frac{16E_{J}}{3E_{2e/3}}\right]\frac{E_{2e/3}}{8E_{J}}\right).
\end{equation}
As a result, the tunneling amplitude for the topological junction is given by~\cite{Supplementary} 
\begin{equation}
\nu_{6\pi}=\sqrt{3}\frac{4E_c}{\sqrt{\pi}}\left(\frac{2E_J}{E_c}\right)^{3/4}e^{-S_{6\pi}}.
\end{equation}
The tunneling amplitude $\nu_{6\pi}$ is sufficiently smaller than $\nu_0$ due to a factor of $3$ in the exponent ($S_0\gg1$).

If we now consider a system consisting of such a junction with a large shunting resistance (underdamped junction), $R>R_Q=2\pi/(2e)^2$, and apply a current, the junction would be in an effectively insulating regime up to some maximal value of the applied current $I_m$, determined by the dispersion of the lowest band~\cite{Likharev1985,Ioffe2007}, which can be seen as a sharp voltage peak $V_m=RI_m$. The value of this current depends on the band width $4\nu_{6\pi}$ and is given by~\cite{Supplementary}
\begin{equation}\label{eq:ic6pi}
I_m=e96\sqrt{3\pi}E_{c}\left(\frac{2E_{J}}{E_{c}}\right)^{3/4}e^{-S_{6\pi}}\frac{R_{Q}}{R}.
\end{equation}
\begin{figure}
\includegraphics[width=6.5cm]{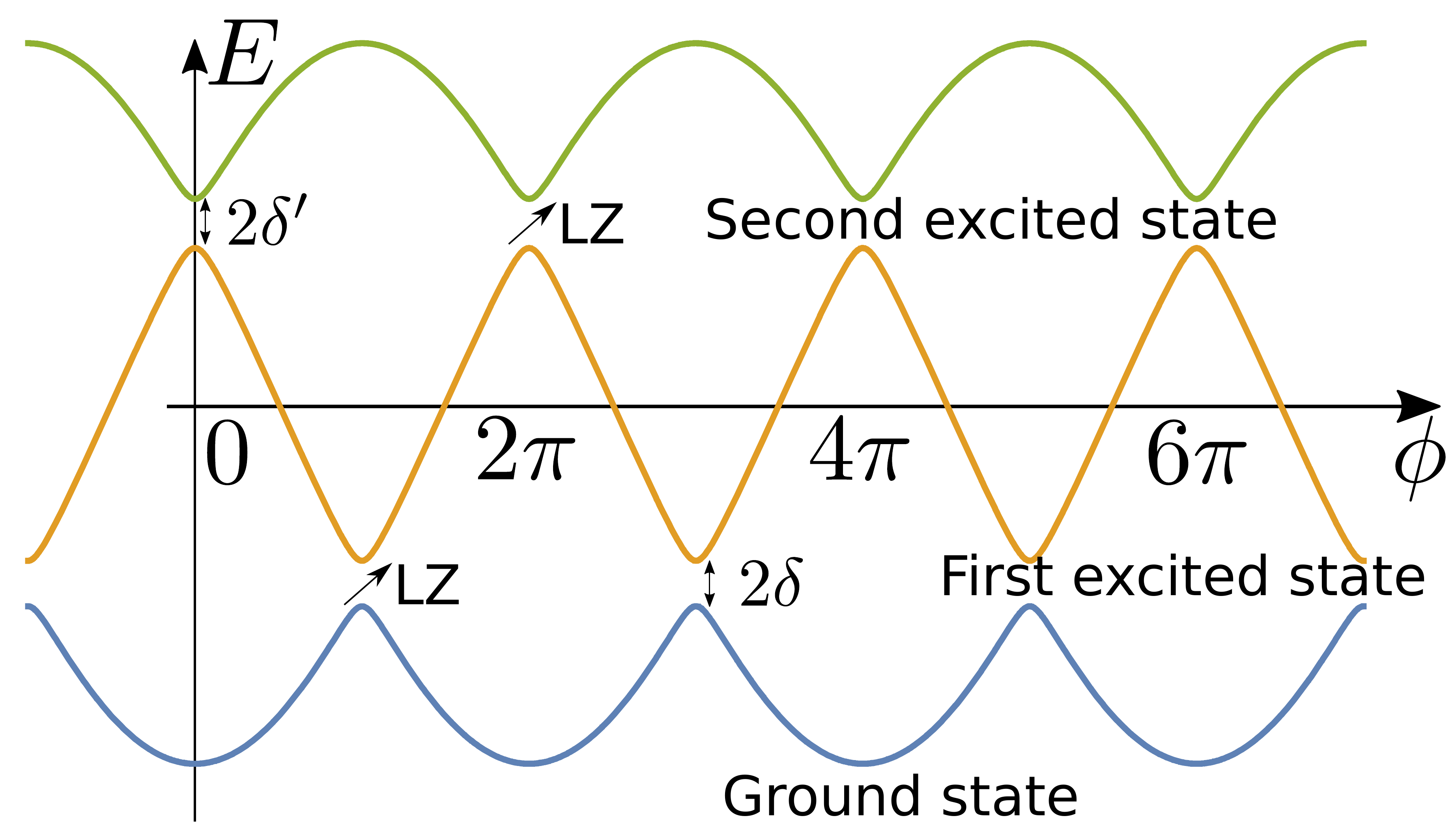}
\caption{\label{fig:6pi_splitting}
The spectrum of a three-level system, formed by a pair of localized $\mathbb{Z}_3$ parafermions with degeneracy lifting due to finite size effects of overlapping parafermions.}
\end{figure}

Finite-size effects may lift the degeneracy between ground states and result in transitions between the three possible states of the parafermion pair on the junction. In particular, the overlap with  parafermions localized on the outer sides of the topological system~\cite{Burnell2016} plays a crucial role. A similar effect has been discussed before for a JJ hosting 
MBSs~\cite{Pikulin2012,Dominguez2012}. The resulting spectrum of a three-level system formed by a pair of parafermions localized on the sides of the junction has avoided level crossings at $\pi{n}$: with energy splitting $2\delta$ [at $\pi(2n+1)$] and $2\delta^\prime$ [at $2\pi{n}$], see Fig.~\ref{fig:6pi_splitting}. Away from the avoided level crossings each branch consists of one of the three states with energy $-E_{2e/3}\cos\left([\phi-2\pi{m}]/3\right)$, where $m$ labels the state; at avoided level crossings the state is given by asuperposition of two states with different $m$. 
If 
$\delta$ is small ($\delta\ll{E}_{2e/3}$) and can be treated perturbatively, the ground state energy is given by  
\begin{equation}
E_g\approx\mathrm{min}_{m}\left\{-E_{2e/3}\cos\left(\frac{\phi-2\pi m}{3}\right)\right\}.
\end{equation}
As we consider $\delta\ll{E}_{2e/3}$, we have neglected the corrections to the energy at the avoided crossing points. In the adiabatic limit (discussed in detail below), the phase potential of the topological JJ is given by $U(\phi)\approx-E_J\cos\phi+E_g$, which is $2\pi$ periodic. We can calculate  the instanton action for a $2\pi$ phase slip in a topological junction (it is different from the non-topological action $S_0$ due to the $E_{2e/3}$ term):
\begin{equation}
S_{2\pi}
=S_{0}\left(1+\frac{3}{8}\left[2\mathrm{arcoth}\sqrt{3}-\ln3\right]\frac{E_{2e/3}}{E_{J}}\right).
\end{equation}
The resulting tunneling amplitude takes the form 
\begin{equation}
\nu_{2\pi}=\frac{4E_c}{\sqrt{\pi}}\left(\frac{2E_J}{E_c}\right)^{3/4}e^{-S_{2\pi}}.
\end{equation}
And, finally, we can calculate the maximal value of the current for the insulating regime:
\begin{equation}\label{eq:ic2pi}
I_m=e32\sqrt{\pi}E_{c}\left(\frac{2E_{J}}{E_{c}}\right)^{3/4}e^{-S_{2\pi}}\frac{R_{Q}}{R}.
\end{equation}
As one can see from Fig.~\ref{fig:6pi}, the above analysis is valid, if the phase dynamics may be considered adiabatic in comparison to the dynamics of the state formed by a pair of localized parafermions. That means that as long as we can neglect Landau-Zener transitions (LZT) at $\phi=2\pi(2n+1)$, the effective potential is determined by the ground state energy of the topological junction for the fixed phase $E_g-E_J\cos\phi$. The probability of LZT is given by 
\begin{equation}
P_{LZ}=\exp\left(-\frac{2\pi\delta^{2}}{\dot{\phi}E_{2e/3}}\right)\approx\exp\left(-\frac{\delta^{2}}{\nu_{2\pi}E_{2e/3}}\right).
\end{equation}
Then if $\delta\gg\sqrt{\nu_{2\pi}E_{2e/3}}$, we can neglect LZTs
and assume the potential to be effectively
$2\pi$ periodic. In the limit $\delta\ll\sqrt{3\nu_{6\pi}E_{2e/3}}$ (the factor $3\nu_{6\pi}$ arises from a new characteristic velocity for phase evolution due to $6\pi$ tunneling: $\dot{\phi}=6\pi\nu_{6\pi}$), we come back to the $6\pi$ periodicity and to the result given by Eq.~(\ref{eq:ic6pi}) (in principal, $\delta^\prime$ may be different from $\delta$, however, the difference between them is not essential for such strong conditions).  As a result, if one can control $\delta$, one can switch the system from an effectively $6\pi$ to an effectively $2\pi$ state, which should be observable as a drop in the voltage peak $V_m=RI_m$ and indicate the presence of parafermions in the system. Moreover, as it was shown in~\cite{Burnell2016} and before for systems hosting MBS~\cite{Rainis2013,Marcus2013,Dmytruk2018} the splitting is oscillating around $0$ as a function of the chemical potential and the applied magnetic field. Therefore, the value of the peak $V_m=RI_m$ changes between two exponentially different values, given by Eqs.~(\ref{eq:ic6pi}) and~(\ref{eq:ic2pi}), if one varies one of this parameters. 

{\it $\mathbb{Z}_4$ case.} The above analysis can also be performed for $\mathbb{Z}_4$ parafermions. A pair of $\mathbb{Z}_4$ parafermions localized on the sides of a junction results in the phase potential~\cite{Cheng2012,Stern2012,Kane2014,Oppen2016}  
\begin{equation}
U=-E_J\cos\phi-\sum_{n=1}^2E_{e/n}\cos\left(\frac{\phi-2\pi{m}}{2n}\right).
\end{equation}
$E_e$ represents  single-electron tunneling, $E_{e/2}$ stands for $e/2$ fractional quasiparticles tunneling, and $m\in\{0,1,2,3\}$ indicates one of the four possible states of the parafermion pair; $E_J$ is a trivial Josephson energy.  In several theoretical works~\cite{Oppen2016,Schmidt2017}, the Cooper-pair tunneling was predicted to be dominating, i.e. $E_J\gg{E}_e, {E}_{e/2}$.  
The harmonic frequency, determining the lowest energy bands, is given by
$
\omega_0\approx\sqrt{8E_JE_c}\left(1+\frac{E_e}{8E_J}+\frac{E_{e/2}}{32E_J}\right).
$
With the assumptions taken above, we calculate the instanton action for tunneling between the lowest minima of the phase potential (expansion in $E_e/E_J$ and $E_{e/2}/E_J$):
\begin{multline}
S_{8\pi}
=4S_{0}\left(1+\frac{1}{8}\left(1+\ln\frac{16E_{J}}{E_{e}}\right)\frac{E_{e}}{E_{J}}\right.\\
+\left.\frac{1}{8}\left(1+\ln\frac{2^{9/2}E_J}{E_{e/2}}\right)\frac{E_{e/2}}{E_{J}}
\right).
\end{multline}
 As a result, we can derive the current $I_m$ at which the junction switches from insulating to conducting state:
\begin{equation}\label{eq:8pi_ic}
I_m=e256\sqrt{\pi}E_{c} \left(\frac{2E_{J}}{E_{c}}\right)^{3/4}e^{-S_{8\pi}}\frac{R_{Q}}{R}.
\end{equation}

\begin{figure}
\includegraphics[width=8.5cm]{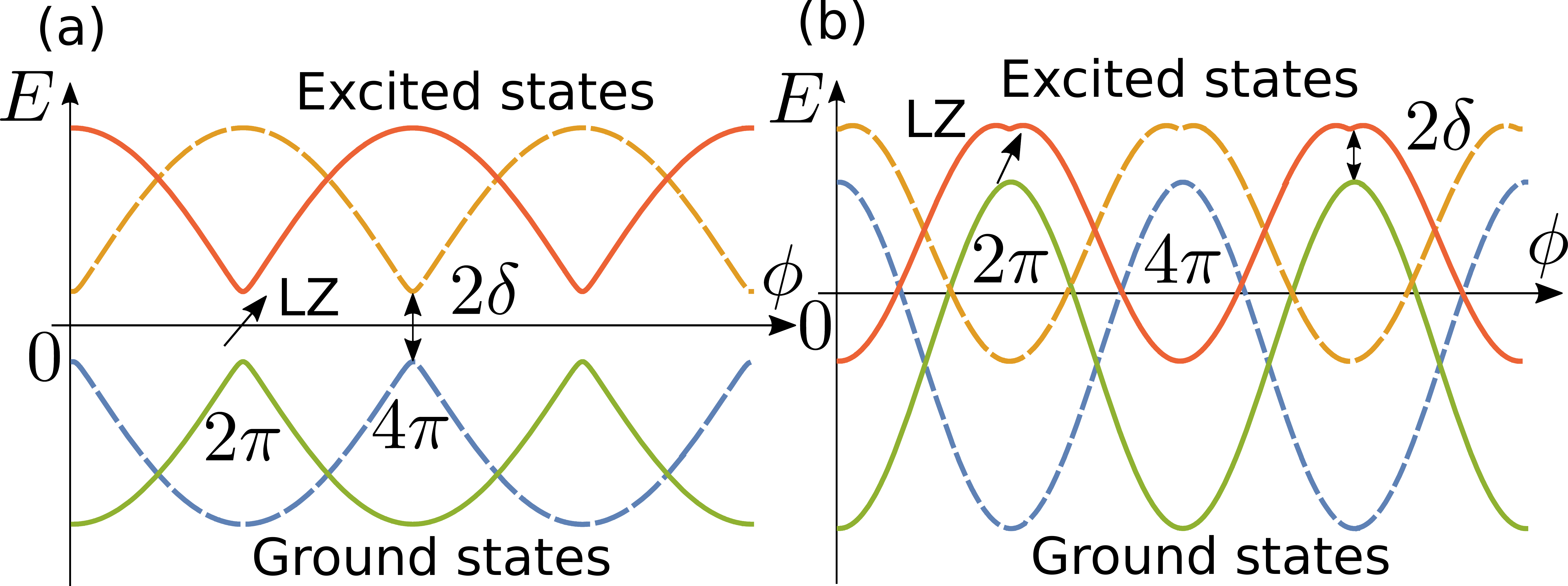}
\caption{\label{fig:8pi}
 The spectrum of a system with TRS formed by a pair of localized $\mathbb{Z}_4$ parafermions: a) with $E_e=0$ and b) with $E_e=2E_{e/2}$. Solid and dashed lines correspond to
states with opposite fermion parity. The Kramers degeneracy at $2\pi{n}$ is lifted due to TRS breaking, while the rest of the crossings survive, being protected by fermion parity. As a result, the ground state is either given by the blue or green branch.}
\end{figure}

Finite-size effects play exactly the same role as in the case of $\mathbb{Z}_3$ parafermions. By varying an applied magnetic field or by shifting the chemical potential, one would tune the overlap with parafermions on the outer edges of the system~\cite{Burnell2016}, which can drive the system to effectively $2\pi$ periodic state with the result similar to one obtained for the non-topological junction, see Eq.~(\ref{eq:ic2pi}) (with additional parametrically small corrections in the tunneling action). However, it is also possible to get a more sophisticated phase periodicity reduction. The systems hosting $\mathbb{Z}_4$ usually posess the time-reversal symmetry (TRS)~\cite{Kane2014,Oppen2016}. If one applies local magnetic fields, the TRS is broken, which would result in lifting of Kramers degeneracy. We can consider the splitting $\delta$ to be small in comparison to the energy scales $E_e$ and $E_{e/2}$. Then, for a fixed phase the energy ground state, formed by a pair of $\mathbb{Z}_4$ parafermions, is given by (green branch in Fig.~\ref{fig:8pi})
\begin{multline}
E_g=-E_{e}\cos\left(\phi/2\right)-\sqrt{\delta^2+E^2_{e/2}\cos^2\frac{\phi}{4}}\\
\approx-E_{e}\cos\left(\phi/2\right)-E_{e/2}\mathrm{max}_m\cos\frac{\phi-4\pi{m}}{4}.
\end{multline}
One should note that only Kramers degeneracies at $2\pi{n}$ are lifted due breaking TRS, all the other crossings remain, as they are protected by fermion parity conservation~\cite{Kane2014,Oppen2016}. If LZT can be neglected (the exact condition is discussed below), the phase potential of the JJ is $U=-E_{J}\cos\phi+E_g$, which allows us to calculate the instanton action for tunneling between the lowest minima:
\begin{multline}
S_{4\pi}=2S_{0}\left(1+\frac{1}{8}\left(1+\ln\frac{16E_{J}}{E_{e}}\right)\frac{E_{e}}{E_{J}}\right.\\
\left.+\frac{1}{16}\left(1+\ln\frac{8E_J}{E_{2/3}}\right)\frac{E_{e/2}}{E_{J}}\right).
\end{multline}
As a result, we can determine the critical current for the insulating regime 
\begin{equation}
I_m=e64\sqrt{2\pi}E_{c} \left(\frac{2E_{J}}{E_{c}}\right)^{3/4}e^{-S_{4\pi}}\frac{R_{Q}}{R}.
\end{equation}
As long as $\delta\gg\sqrt{2\nu_{4\pi}E_{e/2}}$ (negligible LZT), the above assumption is valid. While in the limit $\delta\ll\sqrt{4\nu_{8\pi}E_{e/2}}$ the LZT probability is almost $1$, which allows us to treat the phase potential as effectively $8\pi$ periodic [and reproduce the results derived above without degeneracy lifting, Eq.~(\ref{eq:8pi_ic})].  

{\it Discussion and conclusions.} The above analysis provides a promising method to establish the presence of parafermions in systems that are expected to support these exotic topological bound states. In this work we do not specify the mechanism responsible for creation of parafermions, as there are numerous different approaches~\cite{Fendley2012,Kane2014,Yakoby2014,Loss2014,Fisher2014,Stern2014,Schmidt2015,Klinovaja2014,Klinovaja2015,
Fendley2016,Beri2016,Oreg2017,Schmidt2017,Alicea2018,Wu2018,Mora2018,Laubscher2019,Trauzettel2019,Kim2020}, all of which still require experimental verification. 
The method consists of measuring $I$-$V$ characteristics of the current-biased junction in an underdamped regime at different values of splitting $\delta$ at avoided crossings. As it was shown in~\cite{Burnell2016}, the splitting due to finite-size effect is oscillating around zero as a function of the chemical potential and magnetic field (similar to junctions supporting MBSs~\cite{Pikulin2012,Dominguez2012}). As a result, if one of these parameters is varied, the system oscillates between the regimes of low and high LZT probabilities with significantly different values of the peak $V_m=RI_m$ (due to different effective periodicity of the phase potential). Moreover, for systems with TRS (no applied magnetic fields) one can switch to a state with reduced periodicity applying local magnetic field.  The  results obtained here may be easily generalized to  systems hosting $\mathbb{Z}_n$ parafermions with any integer $n$. The voltage peak should be at
 \begin{equation}
 I_m=e32\sqrt{\pi n^3/l^3}E_{c} \left(\frac{2E_{J}}{E_{c}}\right)^{3/4}e^{-S_{2\pi{n}/m}}\frac{R_Q}{R},
 \end{equation}
where  $l<n$ is the reduced periodicity factor arising from finite-size effects or TRS breaking.
The generalized formula is valid as long as the Cooper-pair tunneling is dominating over any fractional quasiparticle tunneling. The tunneling action is given by $S_{2\pi{n}/l}=nS_0/l+...$, where the correction is determined by the terms corresponding to fractional quasiparticle tunneling. Thus, $I_m$ changes significantly if $l$ goes from $l=1$ (negligible splitting) to $l>1$. This non-monotonic behaviour of $I_m$ is specific only for topological junctions, which provides a straightforward way to distinguish a junction hosting Majorana fermions~\cite{Svetogorov2020} or parafermions.

We also have to mention that experimentally it may be difficult to prepare the system in a state  corresponding exclusively to the lowest band. Some population in higher bands would increase the value of $I_m$. However, $I_m$ would still remain exponentially suppressed (see the analysis for $\mathbb{Z}_3$ in~\cite{Supplementary}), therefore, the oscillatory behaviour of $I_m$ as a function of splitting remains. Another important issue to mention is that in systems with intrinsic spin-orbit coupling, which are typically proposed to fabricate a junction supporting parafermions, the real phase dependence of Andreev levels is not exactly given by a cosine~\cite{Lutchyn2012,Moler2015,Park2017,Glazman2017}. However, the periodicity remains the same, which allow us to claim that qualitatively the results remain the same - the higher the effective periodicity of the system is, the narrower are the lowest bands, therefore, the lower is the voltage peak $V_m=RI_m$. 

We thank Flavio Ronetti and Katharina Laubscher for fruitful discussions. This work was supported by the Swiss National Science Foundation and NCCR QSIT.
This project received funding from the European Union's Horizon 2020 research and innovation program (ERC Starting Grant, grant agreement No 757725).
\bibliographystyle{apsrev4-1}
\bibliography{Para}
\end{document}


\title{Supplemental Material:\\
"Insulating regime of an underdamped current-biased Josephson junction\\ supporting $\mathbb{Z}_3$ and $\mathbb{Z}_4$ parafermions"}
\author{Aleksandr E. Svetogorov}
\author{Daniel Loss}
\author{Jelena Klinovaja}
\affiliation{University of Basel, Department of Physics, Klingelbergstrasse, 4056 Basel, Switzerland}
\maketitle
\section{Critical current $I_m$ for a topological junction supporting $\mathbb{Z}_3$ parafermions}
In this section, we calculate $I_m$ for a topological junction supporting $\mathbb{Z}_3$ parafermions.  To do this we start with calculating the dispersion of the lowest band in the phase potential of the Hamiltonian
\begin{equation}
H=\frac{q^2}{2C}-E_J\cos\phi-E_{2e/3}\cos\frac{\phi}{3},
\end{equation}corresponding to a junction not biased by a current. Here, $C$ is the capacitance of the junction, $q$ is the charge on the junction, $\phi$ is the superconducting phase difference on the junction. The lowest band is given by
\begin{equation}
E^{(0)}\left(k\right)=\frac{\omega_{0}}{2}-2\nu_{6\pi}\cos\left(6\pi k\right),
\end{equation}
where tunneling amplitude $\nu_{6\pi}$ can be expressed through $\nu_0$, which is the $2\pi$ tunneling amplitude in the non-topological case~\cite{ABC,Svetogorov2020}:
\begin{equation}
\nu_{6\pi}=\sqrt{\frac{S_{6\pi}}{2\pi}}{\cal{N}}e^{-S_{6\pi}}\approx\sqrt{\frac{S_{6\pi}}{S_{2\pi}}}e^{-S_{6\pi}+S_{0}}\nu_{0},
\end{equation}
where $\cal{N}$ is determined by the reduced determinant (with excluded zero mode) of an operator that corresponds to
the second variation of the imaginary-time action in an instanton technique approach. As the difference in phase potential in the topological and non-topological ($E_{2e/3}=0$) cases is only parametrically small $O\left(E_{2e/3}/E_J\right)$, $\cal{N}$ can be considered the same for both. Thus, the only principal difference is in the instanton action. The tunneling amplitude for a trivial junction is given by ~\cite{Likharev1985,Matveev2002}
\begin{equation}\label{eq:nu_0}
\nu_0=\frac{4E_c}{\sqrt{\pi}}\left(\frac{2E_J}{E_c}\right)^{3/4}e^{-S_{0}}.
\end{equation} 
The instanton action for the $2\pi$ phase slip is given by $S_{0}=\sqrt{8E_J/E_c}$~\cite{Likharev1985,Matveev2002}.  We can calculate the instanton action for the $6\pi$ phase slip in a topological junction:
\begin{equation}
S_{6\pi}=\intop_{0}^{6\pi}\sqrt{\frac{E_{J}\left(1-\cos\phi\right)+E_{2e/3}\left(1-\cos\frac{\phi}{3}\right)}{4E_{c}}}d\phi
=3S_{0}\left(1+\left[1+\ln\frac{16}{3}-\ln\frac{E_{2e/3}}{E_{J}}\right]\frac{E_{2e/3}}{8E_{J}}+O\left(\frac{E_{2e/3}^2}{E_{J}^2}\right)\right).
\end{equation}
As a result, we obtain the tunneling amplitude giving rise to the lowest band in the effective phase potential (we neglect corrections due to the non-zero $E_{2e/3}$ in the pre-exponential factor) 
\begin{equation}
\nu_{6\pi}\approx\sqrt{3}\frac{4E_c}{\sqrt{\pi}}\left(\frac{2E_J}{E_c}\right)^{3/4}e^{-S_{6\pi}}.
\end{equation}

Now let us consider an insulating regime of an underdamped topological junction biased by a small current $I\ll{I}_m$. Using the analogy of a particle moving in a one-dimensional potential, we can write the semiclassical equations of motion~\cite{Likharev1985,Ioffe2007}: 
\begin{align}
\frac{d\phi}{dt}={}&{}\frac{dE^{(0)}}{dk},\\
\frac{dk}{dt}={}&{}\frac{I}{2e}-\frac{R_Q}{R}\frac{d\phi}{dt},
\end{align}
where $\tilde{q}=2ek$ is the quasi-charge in analogy to quasi-momentum for a particle in a one-dimensional potential: in each band , the energy is periodic in $k$, i.e., $E^{(n)}(k+1)=E^{(n)}(k)$, therefore, we can restrict ourselves to the first Brillouin zone, $-1/2<k<1/2$; $\phi$ is the phase difference; $R\gg{R}_Q$ is the shunting resistance, while $R_Q=1/(2e)^2$ is the resistance quantum.
Then, up to a maximum current $I_m=2e\,\mathrm{max}\{\frac{dE^{(0)}}{dk}\}\frac{R_Q}{R}$, the current $I$ flows through the external resistance $R$ as there is a stationary solution with constant $k$: 
\begin{equation}
\frac{d\phi}{dt}=\frac{I}{2e}\frac{R}{R_Q}.
\end{equation}
This stationary regime corresponds to an effectively insulating state of the junction. At stronger driving currents, i.e., $I>I_m$, there is no longer a solution with constant $k$ and the system enters the regime of Bloch oscillations. In this regime, for low dissipation, the motion is periodic in $k$ \cite{Ioffe2007}.  As a result, the voltage $V$ is decreasing with the increase of the driving current $I$ and the junction is no longer in the insulating state. Now let us write down the expression for $I_m$:
\begin{equation}
I_m=e96\sqrt{3\pi}E_{c}\left(\frac{2E_{J}}{E_{c}}\right)^{3/4}e^{-S_{6\pi}}\frac{R_{Q}}{R}.
\end{equation}

\section{Contribution of higher bands}

\begin{figure}[b]
\includegraphics[scale=0.6]{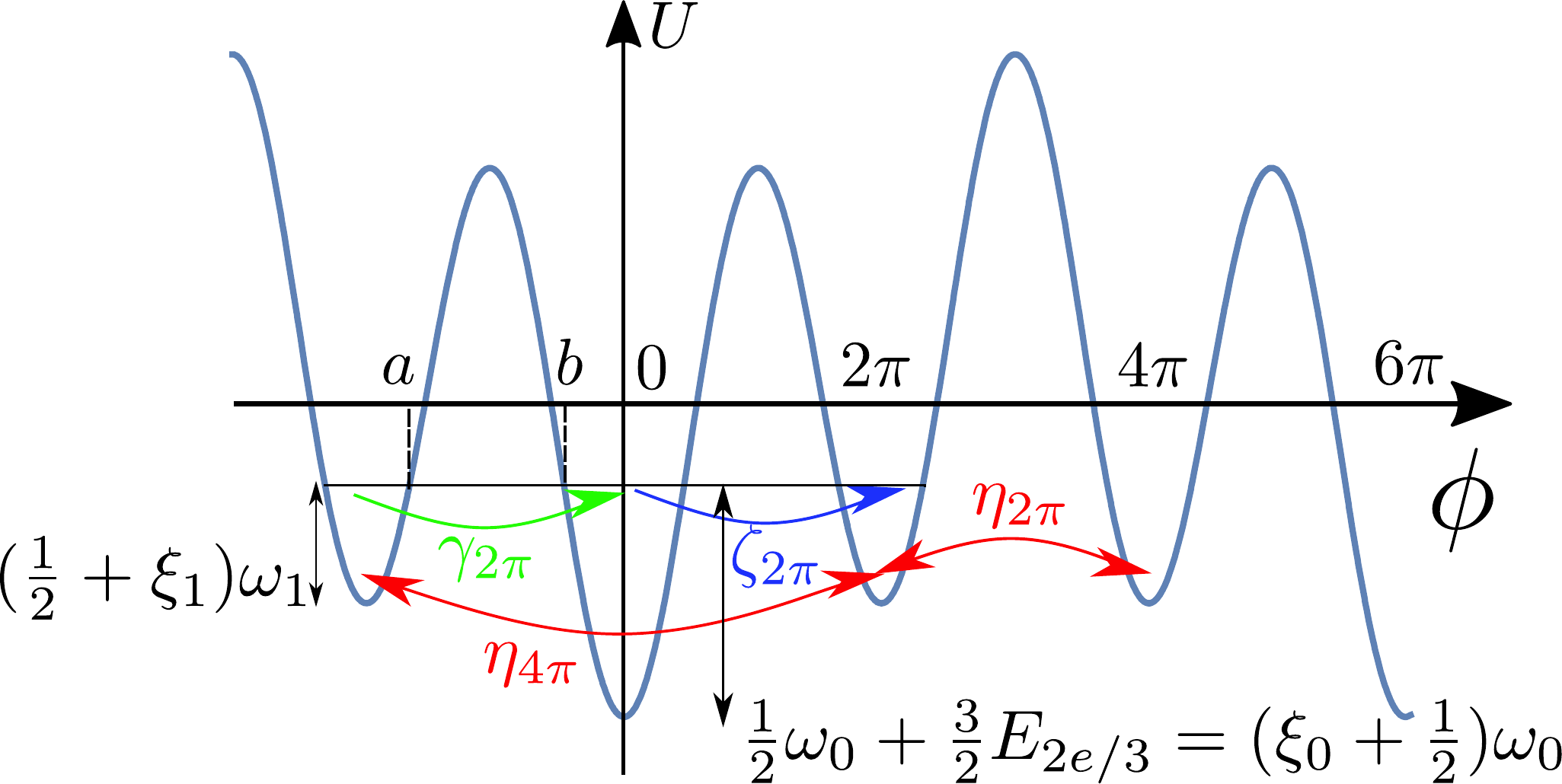}
\caption{\label{fig:tunneling}
Schematic representation of tunneling processes that determine the next higher band. The levels in the local minima can be viewed as the ones slightly shifted by $\xi_0\ll1$ and $\xi_1\ll1$ from the integer-valued harmonic levels.}
\end{figure}
Here we calculate the contribution of a higher energy band to the maximal current $I_m$ for the insulating regime of a Josephson junction supporting $\mathbb{Z}_3$  parafermions. An attempt to prepare the system in the ground state may result in some population in the higher band: $\psi=c_{0}\psi_{0}+c_{1}\psi_{1}$, where $\psi_0$ is the wave function for the lowest band, $\psi_1$ for the next band, $c_0^2+c_1^2=1$. As a result, the critical current for the insulating regime $I_c$ depends not only on the tunneling between the absolute minima of the potential, but also on the tunneling between the higher local minima, see Fig.~\ref{fig:tunneling}. Therefore, the maximal current for an insulating regime of the underdamped junction is given by
\begin{equation}\label{eq:im}
I_{m}=24\pi e\frac{R_{Q}}{R}\left(c_{0}^{2}\nu_{6\pi}+c_{1}^{2}\eta_{6\pi}\right).
\end{equation}
The higher band dispersion takes the form
\begin{equation}
E^{\left(1\right)}\left(k\right)=\frac{1}{2}\omega_{1}+\frac{3}{2}E_{2e/3}-2\eta_{6\pi}\cos\left(6\pi k\right),
\end{equation}
where $\eta_{6\pi}$ is the tunneling amplitude between the ground state in the local minimum at $\phi\approx-2\pi$ and local minimum at $\phi\approx4\pi$. This tunneling process consists of approximately $4\pi$ and $2\pi$ tunneling processes: $\eta_{6\pi}=\frac{\eta_{4\pi}\eta_{2\pi}}{\eta_{4\pi}+\eta_{2\pi}}$, see Fig.~\ref{fig:tunneling}. One can easily see that $\eta_{4\pi}\ll\eta_{2\pi}$, due to longer tunneling, which results in $\eta_{6\pi}\approx\eta_{4\pi}$.
The tunneling amplitude $\eta_{4\pi}$ between $-2\pi$ and $2\pi$ can be calculated
by connecting the corresponding wave functions in these minima. The phase potential is
\begin{equation}
U=-E_{J}\cos\phi-E_{2e/3}\cos\frac{\phi}{3}.
\end{equation}
As the tunneling passes through the absolute minimum at $\phi=0$, which is lower than the local minima, the instanton technique cannot be used to calculate the tunneling amplitude. Moreover, one cannot use the WKB solution for the low states in the local minima of the potential, therefore simple WKB calculation of the tunneling amplitude also does not work. However, we can write the exact solutions in the minima and connect them through WKB solutions under the barrier, as will be shown in the following.
We can expand the phase potential around the local minimum $\phi_{min}\approx-2\pi$ and introduce a harmonic frequency $\omega_1\approx\sqrt{8E_JE_c}\left(1-\frac{E_{2e/3}}{36E_J}\right)$, correponding to this minimum. The Schr{\"o}dinger equation reads
\begin{equation}
-4E_c\left[\frac{\partial^2}{\partial\phi^2}-\frac{\omega_1^2}{(8E_c)^2}(\phi-\phi_{min})^2\right]\psi_n=E_n\psi_n.
\end{equation}
The energy $E_n$ is now calculated from the bottom of the minimum. As we have used the harmonic approximation, the spectrum is well known and given by $E_n=\left(n+\frac{1}{2}\right)\omega_1$. The exact value of the minimum $\phi_{min}$ is determined by the equation
\begin{equation}
E_J\sin\phi_{min}+\frac{1}{3}E_{2e/3}\sin\frac{\phi_{min}}{3}=0.
\end{equation}
As we look for the minimum close to $-2\pi$, we can expand in $E_{2e/3}/E_J$ and get
\begin{equation}
\phi_{min}=-2\pi+\frac{\sqrt{3}}{6}\frac{E_{2e/3}}{E_J}+O\left(\frac{E^2_{2e/3}}{E^2_J}\right).
\end{equation}
It is convenient to introduce a new variable 
\begin{equation}
x=\sqrt{\frac{\omega_{1}}{8E_{c}}}\left(\phi-\phi_{min}\right),
\end{equation}
and get a dimensionless equation:
\begin{equation}
\psi^{\prime\prime}+\left(2n+1-x^2\right)\psi=0,
\end{equation}
where $\psi^{\prime\prime}=\frac{\partial^2}{\partial{x}^2}\psi$. Introducing $\psi(x)=e^{-x^2/2}\zeta(x)$ we arrive at the equation for Hermite polynomials:
\begin{equation}\label{eq:hermite}
\zeta^{\prime\prime}-2x\zeta^\prime+2n\zeta=0.
\end{equation}
 Now one can ask, why we have even started all this formal discussion, if it was evident from the very beginning that the solution near the bottom of the local minimum should just correspond to the standard harmonic oscillator state. However, this is true only if one neglects the tunneling. The tunneling shifts the levels (strictly speaking, it turns discrete levels into bands, as the phase potential is periodic, however, to calculate the tunneling amplitude, it is enough to consider the tunneling process to the next local minimum $\phi\approx2\pi$). Therefore, what we need is exactly the shift from the non-perturbed harmonic oscillator state. To do so, we can consider a solution with $n$ slightly shifted from an integer value, i.e. the ground state in the local minimum is shifted from $n=0$ to $n=\xi_1\ll1$. As a result we have
 parabolic cylinder functions which can be seen as generalization
of Hermite polynomials to the case of non-integer numbers $n$. A general solution of Eq.~(\ref{eq:hermite}) can be written in the contour-integral form (using the Laplace method for solving linear equations):
\begin{equation}
\zeta(x)=\intop_C\frac{dt}{t^{n+1}}e^{xt-t^2/4}.
\end{equation}
 As we consider the tunneling to be exponentially weak, in the left local minimum we have $n=\xi_{1}\ll1$. 
Now we want to find the asymptotics of the wave function in the local minimum. We have
\begin{equation}
\psi(x)=\frac{1}{2\pi{i}}e^{-x^2/2}\intop_C\frac{dt}{t^{n+1}}e^{xt-t^2/4},
\end{equation}
where the factor $1/(2\pi{i})$ is chosen to simplify the final expression. We can fix the solution with a condition that there is only a decaying exponent at $x\rightarrow-\infty$ (as we consider tunneling to the right, we do not consider any local minima to the left). Therefore, the contour $C$ should not pass through the saddle point, when $x\rightarrow-\infty$. The saddle point is at $t_c=2x$. Then we can choose a branch cut given by $t\in(0,+\infty)$, while the contour $C$ starts from $t=+\infty-i0$, follows the branch cut from below, runs around $t=0$ and then to $t=+\infty+i0$ above the branch cut. Above the branch cut we take $1/t^{n+1}=1/|t|^{n+1}$, while below it the phase $-2i\pi(n+1)$ is acquired. Then the total integral can be seen as a sum of two integrals, $I=I_b+I_a$, where $I_b$ is the integral running below the branch cut, while $I_a$ runs above it. Both integrals do not  pass through the saddle point for $x\rightarrow{-}\infty$. Then we have
\begin{equation}
I_a\approx{e}^{-x^2/2}\frac{1}{2\pi{i}}\intop_{0}^{+\infty}\frac{dt}{t^{n+1}}e^{-|x|t}=|x|^n\Gamma(-n)\frac{e^{-x^2/2}}{2\pi{i}}.
\end{equation}
And for the second integral we have $I_b=-I_ae^{-2i\pi{n}}$. Combining them together and using the fact that $\Gamma(-n)\Gamma(n+1)=-\frac{\pi}{\sin(\pi{n})}$ we get
\begin{equation}
I\left(x\rightarrow-\infty\right)=-\frac{e^{-i\pi{n}}}{\Gamma\left(n+1\right)}|x|^{n}e^{-x^2/2}.
\end{equation}
To find the opposite asymptotic expression for $x\rightarrow{+}\infty$, we can rewrite $I_a$ as
\begin{equation}
I_a=\frac{{e}^{-x^2/2}}{2\pi{i}}\left(\intop_{-\infty}^{+\infty}\frac{dt}{t^{n+1}}e^{xt-t^2/4}-\intop_{-\infty}^{0}\frac{dt}{t^{n+1}}e^{xt-t^2/4}\right)\approx\frac{{e}^{-x^2/2}}{2\pi{i}}\left(\frac{\sqrt{\pi}e^{x^2}}{2^nx^{n+1}}-x^n\Gamma(-n)e^{-i\pi(n+1)}\right).
\end{equation}
As a result, we have
\begin{equation}
I\left(x\rightarrow+\infty\right)=e^{-i\pi{n}}\left[\frac{\sin\pi n}{2^{n}\sqrt{\pi}}\frac{e^{x^2/2}}{x^{n+1}}
-\frac{e^{-i\pi n}}{\Gamma\left(n+1\right)}x^{n}e^{-x^2/2}\right].
\end{equation}
Then, the asymptotic expressions for the normalized wave function in the left local minimum is given by
\begin{align}
&\psi\left(x\rightarrow-\infty\right)=C_L\frac{|x|^{n}e^{-x^2/2}}{\Gamma\left(n+1\right)},\\
&\psi\left(x\rightarrow+\infty\right)=-C_L\left[\frac{\sin\pi n}{2^{n}\sqrt{\pi}}\frac{e^{x^2/2}}{x^{n+1}}
-\frac{e^{-i\pi n}}{\Gamma\left(n+1\right)}x^{n}e^{-x^2/2}\right],
\end{align}
where $C_L$ is the normalization constant.

For the classically forbidden region we can write the WKB wave
function
\begin{equation}
\psi_{WKB}=\frac{A_{L}}{\sqrt{|p|}}e^{-S(x)}+\frac{B_{L}}{\sqrt{|p|}}e^{S(x)}.
\end{equation}
To connect it to the solution in the minimum it is usefull to find
the approximate expression for the WKB function near the left turning point $x_0=\sqrt{\frac{\omega_{1}}{8E_{c}}}\left(a-\phi_{min}\right)$, see Fig.~\ref{fig:tunneling}: 
\begin{equation}
S=\intop_{a}^{\phi}|p\left(\phi'\right)|\,d\phi'=\intop_{x_{0}}^{x}\sqrt{x'^{2}-x_{0}^{2}}\,dx'\approx\frac{x^{2}}{2}+\frac{x_{0}^{2}}{2}\ln\frac{x_{0}}{2\sqrt{e}x},
\end{equation}
resulting in 
\begin{equation}
\psi_{WKB}\approx A_{L}\left(\frac{2\sqrt{e}}{\sqrt{2n+1}}\right)^{n+1/2}x^{n}\,e^{-x^{2}/2}+B_{L}\,x^{-n-1}\left(\frac{\sqrt{2n+1}}{2\sqrt{e}}\right)^{\left(n+1/2\right)}\,e^{x^{2}/2}.
\end{equation}
We can connect the solutions
\begin{equation}
\frac{e^{-i\pi n}}{\Gamma\left(n+1\right)}x^{n}C_{L}=A_{L}\left(\frac{2\sqrt{e}}{\sqrt{2n+1}}\right)^{n+1/2}x^{n},
\end{equation}
resulting in
\begin{equation}
A_{L}=\frac{e^{-i\pi n}}{\Gamma\left(n+1\right)}\left(\frac{2\sqrt{e}}{\sqrt{2n+1}}\right)^{-n-1/2}C_{L},
\end{equation}
and 
\begin{equation}
-\frac{\sin\pi n}{2^{n}\sqrt{\pi}}\frac{C_{L}}{x^{n+1}}=B_{L}\,x^{-n-1}\left(\frac{\sqrt{2n+1}}{2\sqrt{e}}\right)^{\left(n+1/2\right)},
\end{equation}
resulting in
\begin{equation}
B_{L}=-\sqrt{2}\frac{\sin\pi n}{\sqrt{\pi}}\left(\frac{\sqrt{2n+1}}{\sqrt{e}}\right)^{-\left(n+1/2\right)}C_{L}.
\end{equation}
Using the fact that $n=\xi_{1}\ll1$, we can write
\begin{equation}
A_{L}\approx\left(2\sqrt{e}\right)^{-1/2}C_{L},\quad B_{L}\approx-\sqrt{2\pi}\xi_{1}e^{1/4}C_{L}.
\end{equation}
The actual $4\pi$ tunneling consists of two $2\pi$ tunnelings $\gamma_{2\pi}$ and $\zeta_{2\pi}$: $\eta_{4\pi}=\frac{\gamma_{2\pi}\zeta_{2\pi}}{\gamma_{2\pi}+\zeta_{2\pi}}$, see Fig.~\ref{fig:tunneling}. The intermediate state in the absolute minimum at $\phi=0$ is shifted from the ground state of the minimum by $3E_{2e/3}/2\ll\omega_p$, which allows us to treat it as a state with $n=\xi_0\ll1$.
On the right-hand side of the barrier (in the absolute minimum at $\phi=0$) the solution can be found
exactly in the same way, just by exchanging the $-\infty\leftrightarrow+\infty$
asymptotics, and introducing $y=\sqrt{\frac{\omega_{0}}{8E_{c}}}\phi$. The harmonic frequency in the minimum is different from the one in the local minimum at $\phi\approx-2\pi$: $\omega_0=\sqrt{8E_c\left(E_J+E_{2e/3}/9\right)}$, however, the difference is parametrically small. 
The WKB solution under the right side of the barrier can be written
in the form
\begin{equation}
\psi_{WKB}=\frac{A_{R}}{\sqrt{|p|}}e^{-S(y)}+\frac{B_{R}}{\sqrt{|p|}}e^{S(y)},
\end{equation}
where the WKB action is calculated from the right turning point $y_{0}=-\sqrt{\frac{\omega_{0}}{8E_{c}}}b$.
We connect solutions under the barrier and in the classical region
on the right exactly in the same way as we have done  for the left
side:
\begin{equation}
A_{R}=\frac{e^{-i\pi n}}{\Gamma\left(n+1\right)}\left(\frac{2\sqrt{e}}{\sqrt{2n+1}}\right)^{-n-1/2}C_{R},\text{ }B_{R}=-\sqrt{2}\frac{\sin\pi n}{\sqrt{\pi}}\left(\frac{\sqrt{2n+1}}{\sqrt{e}}\right)^{-\left(n+1/2\right)}C_{R}.
\end{equation}
Now we can use the fact that on the right side we also have $n=\xi_{0}\ll1$ (shift from the ground state with $n=0$):
\begin{equation}
A_{R}\approx\left(2\sqrt{e}\right)^{-1/2}C_{R},\quad B_{R}\approx-\sqrt{2\pi}\xi_{0}e^{1/4}C_{R}.
\end{equation}
Connecting the WKB solution under the barrier is simple, as the actions
in the exponents are calculated from the turning points, therefore
\begin{equation}
B_{R}=A_{L}e^{-S_\eta},\quad\,A_{R}=B_{L}e^{S_\eta},
\end{equation}
where $S_\eta$ is the tunneling action between the turning points.
As a result, we finally get
\begin{equation}
-\sqrt{2\pi}\xi_{0}e^{1/4}C_{R}=\left(2\sqrt{e}\right)^{-1/2}C_{L}e^{-S_\eta},-\sqrt{2\pi}\xi_{1}e^{1/4}C_{L}=\left(2\sqrt{e}\right)^{-1/2}C_{R}e^{-S_\eta},
\end{equation}
which yields 
\begin{equation}
\xi_{1}\xi_{0}=\frac{e^{-2S_\eta}}{4\pi e}.
\end{equation}
We should take into account the condition on the energy
\begin{equation}
\omega_{0}\xi_{0}\approx\omega_{1}\xi_{1}+3E_{2e/3}/2,
\end{equation}
from which we get
\begin{equation}
\frac{e^{-2S_\eta}}{4\pi e\xi_{1}}\omega_{0}=\omega_{1}\xi_{1}+3E_{2e/3}/2.
\end{equation}
The tunneling matrix element is $\gamma_{2\pi}=\omega_{1}\xi_{1}$, then
\begin{equation}
\gamma_{2\pi}^{2}+\frac{3}{2}E_{2e/3}\gamma_{2\pi}-\frac{e^{-2S_\eta}}{4\pi e}\omega_{0}\omega_{1}=0,
\end{equation}
which gives
\begin{equation}
\gamma_{2\pi}\approx\frac{\omega_{0}\omega_{1}e^{-2S_\eta}}{6\pi eE_{2e/3}}\approx\frac{4E_{J}E_{c}e^{-2S_\eta}}{3\pi eE_{2e/3}}.
\end{equation}
It is easy to see that for tunneling outside of the absolute minimum $\zeta_{2\pi}=\omega_{0}\xi_{0}\gg\gamma_{2\pi}$, which perfectly fits our intuition, as the quasistate in the absolute minimum is not really occupied, but rather effectively serves as a tunneling channel: 
\begin{equation}
\zeta_{2\pi}=\gamma_{2\pi}+3E_{2e/3}/2\approx3E_{2e/3}/2\gg\gamma_{2\pi}.
\end{equation}
Therefore, $\eta_{6\pi}\approx\gamma_{4\pi}\approx\gamma_{2\pi}$.
Now we need to calculate the tunneling action $S_\eta$ between the turning points $a$ and $b$:
\begin{equation}
S_\eta=\frac{1}{2}\sqrt{\frac{E_{J}}{E_{c}}}\intop_{a}^{b}\sqrt{\cos a-\cos x+\frac{E_{2e/3}}{E_{J}}\left(\cos\frac{a}{3}-\cos\frac{x}{3}\right)}dx
=\sqrt{\frac{8E_J}{E_c}}\left[1+O\left(\frac{E_{2e/3}}{E_J}\right)+O\left(\sqrt{\frac{E_c}{E_J}}\right)\right].
\end{equation}
Then the total amplitude, determining the band, is
\begin{equation}
\eta_{6\pi}\sim{e}^{-2\sqrt{\frac{8E_J}{E_c}}}\gg\nu_{6\pi}.
\end{equation}
Thus, if the population of the higher band $c_1$ is not negligible in Eq.~(\ref{eq:im}), the result is mostly determined by $\eta_{6\pi}\gg\nu_{6\pi}$. However, as $\eta_{6\pi}\ll\nu_{2\pi}\sim\exp\left(-\sqrt{8E_J/E_c}\right)$, the resulting $I_m$ is still sufficiently different from the case with significant splitting $\delta$.

\bibliographystyle{apsrev4-1}
\bibliography{Para}